
\input phyzzx
\def\nl{ \hfill\break }     
\def\w{\wedge}               %
\def\half{{\textstyle{1\over 2}}}    
\def\quart{{\textstyle{1\over 4}}}  
\def\3{\,{^3}\!}                
\def\e{{\tilde e}}
\def\N{\ N\!\!\!\!\!_{_\sim}\  } 
\def\n{ n\!\!\!\!_{_\sim}\ }
\def\m{ m\!\!\!\!_{_\sim}\ }
\pubtype{}
\nopubblock
\date{October 1991}
\titlepage
\title{\bf The  reality conditions for the new canonical variables of General
Relativity }
\author{Giorgio Immirzi}
\address{ Dipartimento di Fisica, Universit\'a di Perugia, and I.N.F.N.,
sez.  di Perugia.\foot{e--mail: immirzi@perugia.infn.it} }
\abstract
We examine the constraints and the reality conditions that have to be
imposed  in the canonical theory of 4--d gravity formulated in terms of
Ashtekar variables. We find that the polynomial reality conditions are
consistent with the constraints, and make the theory equivalent to
Einstein's, as long as the inverse metric is not degenerate; when it is
degenerate, reality conditions cannot be consistently imposed in general,
and the theory describes complex general relativity.
\vfil
\sequentialequations
\chapter{Introduction.\hfill\hfill}
\REF\asht{A. Ashtekar,   Phys. Rev. Lett.  57 (1986) 2244; Phys. Rev.
 D36 (1986) 1587;  {\it New perspectives in canonical gravity} (with
invited contributions),  Bibliopolis, Napoli1988.}
\REF\ADM{ R. Arnowitt, S. Deser, C.W. Misner,  in {\it Gravitation: an
introduction to current research}, edited by L. Witten (Wiley 1962).}
\REF\carlo{C. Rovelli,  Class. and Quantum Gravity  }
\REF\reggedauria{R. D'Auria, T. Regge, Nucl. Phys. B195 (1982) 308}
\REF\wittena{E. Witten, Nucl. Phys. B311 (1988) 46  }
\REF\horo{G. T. Horowitz,  Class. and Quantum Gravity  8 (1991) 587}
\REF\bengta{I. Bengtsson, Class. and Quantum Gravity 8 (1991) 1847}
\REF\ART{ A. Ashtekar, J. Romano, R. Tate,   Phys. Rev. D40 (1989) 2572.}
\REF\Bala{A. Ashtekar, A.P. Balachandran, S. Jo, Int.Jour. of Mod.Phys.
           A4 (1989) 1493}
In 1986, A. Ashtekar found a new canonical formulation of Einstein's
theory of gravity [\asht ], quite different and in many ways much more
appealing than the one that had been evolved by P.A.M. Dirac, P. Bergmann
and R. Arnowitt, S. Deser and C.W. Misner [\ADM ](amongst others) many
years before.  This remarkable work has given a new impulse to
the field of canonical gravity, and has generated a totally new approach
to the problem of formulating a quantum theory [\carlo ].

 In this new formulation the basic canonical variables are a
(densitized) inverse dreibein $\e^{a\mu}$, and a complex gauge field
$A^a_\mu$, and the constraints of the theory are polynomial in these new
variables. In particular, this implies that the equations of the theory are
regular  regardless of whether $\e^{a\mu}$ is invertible or not.
This extension of Einstein's theory has attracted a lot of
attention , especially because it is expected to
play a key role in quantum gravity [\carlo -\bengta ].

A degenerate $\e^{a\mu}$ signals a change of signature of the metric,
which may  be associated with a change in topology; we can hardly expect
our interpretation of the variables of the theory to survive unscathed.
In fact, besides constraints and equations of motion, there are reality
conditions that must be imposed on the variables, i.e. a real section has to
be identified on the constraint surface, for the theory  to describe real
space--time. In this paper we investigate whether these reality conditions
can be consistently imposed on configurations with degenerate
$\e^{a\mu}$, and conclude that in general they cannot.

The reality  conditions do not
descend from the action, like constraints (be they first or second class), but
have to be imposed "by hand" on the initial condition;  all is well if
the time evolution of the system preserves them, i.e. if the Poisson bracket
of the Hamiltonian with a quantity stated  to be real can be seen to be real.

 What we find is that although constraints and reality conditions are
polynomial in the basic variables, the  expression of the time derivative
of a reality condition in terms of constraints and of real quantities
is {\it not} polynomial, and requires the 3--metric to be non degenerate to
be valid. Therefore, when the metric is non degenerate we find that the
theory expressed in terms of the new variables with polynomial reality
conditions  is completely equivalent to Einstein's theory. When degeneracy
occurs, we cannot consistently impose the reality conditions, and we have a
complex general relativity theory.

In my opinion this situation is satisfactory, and the problems of
interpretation which are left open are best investigated looking first at
particular examples.

Before discussing the reality conditions in \S 4, I review the use of
selfdual variables and the derivation the Ashtekar action in \S 2,
and  the structure of the constraints  in \S 3.
I limit myself to the case of pure gravity; the inclusion of matter and of a
cosmological term, along the lines of [\ART ],  would complicate the
argument, but presents no new difficulty.
\chapter{The Ashtekar action.   \hfill\hfill}
 One obtains the action expressed  in terms of Ashtekar
variables by projecting  a Palatini-like action, that
depends only on the self-dual part of the connection
and on the vierbein forms $e^i=e^i_\mu dx^\mu$\foot{
 the Lorentz metric is
$\eta_{ij}=(-+++)$, the space-time metric is
$g_{\mu\nu}=\eta_{ij}e^i_\mu e^i_\nu$,  $E:=\det (e^i_{\,\mu})$,
$\epsilon_{0123}=1$, and I have set $8\pi G=c=1$},
on a space slice $\Sigma_t$ [\Bala ].

Let us begin with a digression on selfdual tensors. Given a real
antisymmetric tensor $A^{ij}$, the complex tensor:
$$A^{+ij}:=\half (A^{ij}-i\half \epsilon^{ij}_{\ \ kl}\,A^{kl})
   \eqn\sii$$
is selfdual, i.e. $\epsilon^{ij}_{\ \ kl}\,A^{+kl} =
i A^{+ij}$.   A selfdual  tensor has $3$ (complex) independent components,
which transform under the $(1,0)$ representation of the Lorentz group.
This can be made explicit using an appropriate Clebsch-Gordon
coefficient to go to a $(1,0)$ basis:
$$A^{+a}:= C^a_{ij}A^{ij} = -\half \epsilon_{abc}A^{bc} + i A^{0a}
  = C^a_{ij}A^{+ij}      \eqn\siii$$
with $a,b,\ldots =1,2,3$. Lorentz transformations are represented by
complex orthogonal matrices ${\cal D}^{(1,0)}(\Lambda)$, unless $\Lambda$
is an ordinary rotation, in which case  ${\cal D}^{(1,0)}_{ab}(\Lambda )=
\Lambda_{ab}$.

It is important to stress that $A^{+a}$ contains  the same information
of $A^{ij}$; for example, for the electromagnetic field tensor $F^{ij}$
we have $F^{+a} = -B_a+iE_a$.

Applied to the  connection form $\omega^{ij}$ this projection gives the
forms $\omega^{+a}$ which, under  infinitesimal local Lorentz
transformations $\Lambda^i_{\ \ j}=\delta^i_j +\lambda^i_{\ \ j}+\ldots $,
transform like:
$$\omega^{+a}_\mu \ \rightarrow\quad
 \omega^{+a}_\mu - (\delta_{ac}\partial_\mu +
\epsilon_{abc}\omega^{+b}_\mu )C^c_{ij}\lambda^{ij}+\ldots :=
\omega^{+a}_\mu -D_\mu (C^a_{ij}\lambda^{ij})+\ldots  \eqn\siv$$
A little more algebra shows that the corresponding curvature form
satisfies:
$$
F^{+ a} :=
 d\omega^{+ a}+\half \epsilon_{abc}\omega^{+ b}\w\omega^{+ c}
  =\half (\partial_\mu \omega^{+ a}_\nu-
        \partial_\nu \omega^{+ a}_\mu+
     \epsilon_{abc}\omega^{+ b}_\mu\omega^{+ c}_\nu )
     dx^\mu\w dx^\nu = C^{ a}_{ij}F^{ij}            \eqn\sv$$
The same arguments apply to the antiselfdual connection $\omega^{-a}=
C^{ij*}_a\omega^{ij}$, and one finds:
$$\omega^{ij}=C_a^{ij}\omega^{+a}+
C_a^{ij*}\omega^{-a}\ ;\qquad F^{ij}=C_a^{ij}F^{+a}+
C_a^{ij*}F^{-a}     \eqn\svi  $$
\REF\samuel{ J. Samuel, Pramana J. of Phys.  28 (1987) L429.}
\REF\JSi{ T. Jacobson, L. Smolin, Class. Quantum Grav.  5 (1988) 583.}
\REF\palatini{A. Palatini, Rend. Circ. Mat. Palermo  43 (1919) 203  }
An action that does use exclusively the self dual components has been
introduced in [\samuel ][\JSi ][\ART ]; it can be written in a variety of ways:
$$
S_A := \int_{\cal M}\half\epsilon_{ijkl}e^i\w e^j\w F^{+kl}
 =i\int_{\cal M}C^a_{ij}e^i\w e^j\w F^+_a =
 \int_{\cal M}{1\over 4}\epsilon_{ijkl}e^i\w e^j\w F^{kl}
+{i\over 2}e^i\w e^j\w F_{ij}
  \eqn\svii  $$
The real part of $S_A$ is the Palatini action [\palatini ], which
is stationary for $\omega^{ij}=\Omega^{ij}$, the Levi--Civita connection
\foot{The Levi--Civita connection is related to the metric compatible
derivative by $\Omega^{ij}_\mu=e^i_\nu\nabla_\mu e^{j\nu}$,
its curvature $R_{\mu\nu}^{\ \ \ ij}$ to the Riemann tensor by
$R_{\mu\nu\rho\sigma}=R_{\mu\nu}^{\ \ \ ij}e_{i\rho}e_{j\sigma}$.},
for which it reproduces the Hilbert action.\nl
But, in spite of its appearance, $S_A$  is completely equivalent
to the Palatini action; in fact, setting
$\omega^{+a}= C^a_{ij}\Omega^{ij}+\phi^a$, one finds:
$$iC^a_{ij}e^i\w e^j\w F^+_a\big|_{\omega^+=C\Omega+\phi}=
\quart e^i\w e^j\w R_{ij}+i\half\epsilon_{abc}C^a_{ij}e^i\w e^j\w\phi^b
\w\phi^c+id( C^a_{ij}e^i\w e^j\w\phi^a)
\eqn\sviii $$
Hence varying with respect to $\omega^{+a}$, one finds that $S_A$ is
stationary at $\omega^{+a}=C^a_{ij}\Omega^{ij}$,  where it is real, and
therefore reproduces  the Hilbert action.

 To develop a canonical formalism appropriate to $S_A$, let us
assume that  ${\cal M}$ is foliated into space-like 3-manifolds  $\Sigma_t$,
indexed by a global time function $t(x)$. Mimicking the procedure one adopts
for gauge theories, one chooses the "time gauge", setting
$$e^0_\mu =-n_\mu :=N\partial_\mu t
\eqn\six$$
which still leaves the theory invariant under local $O(3)$ transformations.
The metric induced on $\Sigma_t$ is given by:
$$q_{\mu\nu}=g_{\mu\nu}+n_\mu n_\nu =e^a_\mu e^a_\nu
\eqn\sx  $$
The flow of time is represented by a time-like vector field $t^\mu$, such
that $t^\mu\partial_\mu t=1$, and
$$t^\mu =Nn^\mu + N^\mu          \eqn\sxi $$
The scalar function  $N$ and the space-like vector field $N^\mu$ are  the
"lapse" and "shift", in the terminology of ADM[\ADM ].
In coordinates adapted to the foliation one would have:
$$t^\mu =(1,0,0,0),\quad n_\mu =(-1,0,0,0),\quad
e^{0\mu}=(-{1\over N},{N^\alpha\over N}),\quad
e^{a\mu}=(0,e^{a\alpha}),\ E=Ne  $$
with $ e=\det (e^a_{\,\alpha})$. Replacing eq.s\six\sxi\ in eq.\svii\ one
finds:
$$S_A=\int dt\int_{\Sigma_t}d^3x
(-\half Ne\epsilon_{abc}e^{a\mu}e^{b\nu}F^{+c}_{\mu\nu}+
iN^\mu e^{a\nu}F^{+a}_{\mu\nu}-i e  t^\mu  e^{a\nu}F^{+a}_{\mu\nu})
\eqn\sxii $$
We now introduce the Ashtekar variables $A^a_\mu$ and $\e^{a\mu}$,
which will play the role of "configuration" and "momentum" variables in the
scheme:
 $$A^a_\mu :=q_\mu^{\ \nu}\omega^{+a}_\nu \ ;\qquad
     \e^{a\mu } := e e^{a\mu }  \eqn\sxiii        $$
$\e^{a\mu}$ is a vector density, which, because of the gauge choice, is
tangent to $\Sigma$, i.e. $\e^{a\mu}= q^\mu_{\ \nu}\e^{a\nu}$,  just like
$N^\mu $; the (complex) Ashtekar connection $A^a_\mu$  is a gauge
field  on the 3-space $\Sigma_t$. Under an infinitesimal  local rotation,
with $\lambda^{ab}  = \epsilon_{abc}\lambda_c $, the transformation
laws will be:
$$\e^{a\mu} \rightarrow \e^{a\mu} +\epsilon_{abc}\e^{b\mu}
\lambda_c +\ldots
\ ;\quad
A^a_\mu \rightarrow  A^a_\mu +  \3 D_\mu \lambda_a +\ldots
\eqn\sxv$$
Covariant 3-derivatives will be well defined for objects which belong to
$(j,0)$ representations  of $O(3,1)$, e.g.:
 $$
 \3 D_\mu S_a =
 q_\mu^{\ \nu}\partial_\nu S_a +\epsilon_{abc}A_\mu^b S_c
\eqn\sxiv $$
where $S_a$ belongs to the $(1,0)$\foot{
for a right-handed Dirac spinor $\psi_R=\half (1+\gamma_5)\psi$, which
belongs to the $(\half ,0)$ representation, we would have
$$q_\mu^{\ \nu }D_\nu\psi_R:=
q_\mu^{\ \nu} (\partial_\nu +\half\omega^{ij}_\nu\Sigma_{ij} )\psi_R=
(q_\mu^{\ \nu}\partial_\nu +\half C^a_{ij}\Sigma^{ij}A^a_\mu )\psi_R
$$
because $\Sigma_{ij}=\quart [\gamma_i,\gamma_j]$ and
$\half\epsilon_{ijkl}\Sigma^{kl}=i\gamma_5\Sigma_{ij}$.
This is why the Ashtekar variables where first introduced in a spinor
representation.},
but also, since for rotations ${\cal D}^{(1,0)}_{ab}(\Lambda )=
\Lambda_{ab}$, if by $S_a$ we mean the $1,2,3$ components of an $S_i$
belonging to the $(\half ,\half )$ representation. It is because of this
unusual situation that we loose control on the reality of $\e^{a\mu}$, and
we have to impose reality conditions.

 With the help of eq.\six\ (more properly, of Frobenius theorem)  one can
show that the curvature of  $A^a_\mu $ is the pull back to $\Sigma_t$ of
$F^{+a}_{\mu\nu}$ :
 $$
 \3 F^a_{\mu\nu} :=q_\mu^{\ \rho} q_\nu^{\ \sigma}
    (\partial_\rho A^a_\sigma -\partial_\sigma A^a_\rho +
     \epsilon_{abc}A^b_\rho A^c_\sigma )  =
      q_\mu^{\ \rho} q_\nu^{\ \sigma}  F^{+a}_{\rho\sigma}
      \eqn\sxvi$$
and that:
$$t^\mu e^{a\nu}F^{+a}_{\mu\nu} =
e^{a\nu} {\cal L}_{\bf t} \omega^{+a}_\nu -e^{a\nu}
 [\partial_\nu (t^\mu \omega^{+a}_\mu )
           +\epsilon_{abc}A^b_\nu (t^\mu \omega^{+c}_\mu)] =
e^{a\nu} {\cal L}_{\bf t} A^a_\nu -e^{a\nu}\3 D_\nu \omega^{+a}_t
      \eqn\sxvii$$
where ${\cal L}_{\bf t}$ indicates the Lie derivative with respect to
$t^\mu$, $\omega^{+a}_t:=t^\mu \omega^{+c}_\mu$.
Putting these ingredients together, defining $\N :=N/e$, and with a final
partial integration, we may write the action \sxii\ as a functional of
quantities defined on 3-space:
$$
 S_A =\int dt\int_\Sigma d^3x\{
-i\e^{a\mu} {\cal L}_{\bf t} A^a_\mu +
  iN^\mu \e^{a\nu}\3 F^a_{\mu\nu}
-\half \N\epsilon_{abc} \e^{a\mu}\e^{b\nu}\3 F^c_{\mu\nu}
 -i\,\omega^{+a}_t\3 D_\nu \e^{a\nu}\}
\eqn\sxviii$$
This remarkable expression is the main point of the whole approach;
 one immediately notices that, as we anticipated, it is polynomial in the
basic variables and their derivatives. Of course, it is complex, which
 presumably makes it useless for path integral quantization, and which
might spoil the whole approach if we let spurious solutions creep in.\nl
Notice that that all the indices that appear in \sxviii\ are space like, as
will all indices from now on. So, if one interprets them as component
indices, they range from 1 to 3.
\chapter{The constraints of the theory.\hfill\hfill}
The most plausible interpretation of  the action \sxviii\ is that it refers to
a phase space described by two sets of {\it complex} variables,  with the
basic Poisson bracket:
$$\{ \e^{a\mu}(x),A^b_\nu (y)\} =
i\delta^{ab}\delta^\mu_\nu \delta^{(3)}(x-y)    \eqn\sxxi$$
 There are also  Lagrange multipliers $\N ,N^\mu $ (real), $ \omega^{+a}_t$
 (complex); if we vary the action with respect to them we find:
$$\eqalign{
(i&)\qquad \tilde H := \epsilon_{abc}
\e^{a\mu}\e^{b\nu}\3 F^c_{\mu\nu} =0 \cr
(ii&)\qquad \tilde F_\mu :=i\e^{a\nu}\3 F^a_{\mu\nu} =0 \cr
 (iii&)\qquad \tilde G^a:=\3 D_\mu \e^{a\mu}=0   \cr}
   \eqn\sxxii$$
 respectively the "scalar", the "vector", and the "Gauss law" constraint.
The constraints reflect the local invariance of the theory, and
are all first class, i.e. their Poisson brackets turn out to
be linear combinations of themselves. To see this we shall follow the
procedure of ref.[\ART ], using  complex test functions
$\lambda_a(x)$, a  real test 3-vector  $f^\mu (x)$, and a real test
density $\n (x)$ to smear the constraints:
$$\eqalign{
  G_\lambda &:= i\int_\Sigma d^3x\lambda_a\3 D_\mu
\e^{a\mu}\cr
 F_{\bf f} &:= i\int_\Sigma d^3x f^\mu (\e^{a\nu}\3F^a_{\nu\mu} +
                        A^a_\mu \3 D_\nu \e^{a\nu})  \cr
 H_n &:= \half \int_\Sigma d^3x \n \epsilon_{abc}
\e^{a\mu}\e^{b\nu}\3 F^c_{\mu\nu}   \cr}
\eqn\sxxiii    $$
 In this notation we may write \sxviii\ in the form:
$$ S_A=
\int dt\int_{\Sigma_t}( -i\e^{a\mu} {\cal L}_t A^a_\mu -{\cal H})\ ;\qquad
 {\cal H} := F_{\bf N} + H_N+G_\omega\ ,\ {\rm with}\ \
\omega^a := \omega^{+a}_t-N^\mu A^a_\mu  \eqn\sxxv$$
 For  the Poisson bracket with the  canonical
variables we use eq.\sxxi\ to find:
 $$
\{ G_\lambda , \e^{a\mu}\} = \epsilon_{abc}\e^{b\mu}
\lambda_c  \ ; \quad
\{ G_\lambda , A^a_\mu \} = \3 D_\mu \lambda_a
\eqn\sxxvi$$
$$\eqalign{
\{ F_{\bf f} , \e^{a\mu} \}  &= f^\nu \partial_\nu\e^{b\mu}
-\e^{a\nu} \partial_\nu f^\mu +\e^{a\mu}\partial_\nu f^\nu
= {\cal L}_{\bf f} \e^{a\mu}  \cropen{12pt}
\{ F_{\bf f}, A^a_\mu \} &= f^\nu\partial_\nu A^a_\mu +
A^a_\nu\partial_\mu f^\nu =  {\cal L}_{\bf f} A^a_\mu  \cr}
 \eqn\sxxvii$$
 $$
\{ H_n,\e^{a\mu}\} = i\3 D_\lambda (\n \epsilon_{abc}
\e^{b\lambda}\e^{c\mu} )\ ;\qquad
\{ H_n,A^a_\mu \} =
  i\, \n \epsilon_{abc} \e^{b\nu}\3 F^c_{\mu\nu}
\eqn\sxxviii$$
Comparing eq.\sxxvi\  with eq.\sxv , we see that the smeared
Gauss law is the generator of gauge rotations; and from eq.\sxxvii , that
the peculiar combination of the vector and the Gauss law constraints
was chosen because it generates diffeomorphisms on $\Sigma$.
On the contrary  the scalar constraint, which is related to the
arbitrariness of the time parameter $t$, does not have an obvious
geometric meaning in 3-space.\nl
With these relations we may derive the equations of motion:
$$\eqalign{
 {\cal L}_{\bf t} A^a_\mu &= \{ {\cal H},A^a_\mu \}=N^\nu\3
F^a_{\nu\mu} +
      i \N\epsilon_{abc}\e^{b\nu}\3 F^c_{\mu\nu}
+\3 D_\mu\omega^{+a}_t   \cropen{12pt}
{\cal L}_{\bf t}\e^{a\mu} &=\{ {\cal H},\e^{a\mu} \}
 -\3 D_\nu (N^\mu\e^{a\nu}-N^\nu\e^{a\mu}
      +i\, \N\epsilon_{abc} \e^{b\mu}\e^{c\nu})
      -\epsilon_{abc}\omega^{+b}_t \e^{c\mu}\cr}   \eqn\sxxix $$
We can also calculate all the Poisson brackets between constraints, and
check that they are indeed all first class. The only one that requires  some
algebraic effort is the scalar-scalar one, which gives:
 $$
\{ H_n , H_m \} = i\int_\Sigma d^3x
(\m\partial_\lambda \n    -\n\partial_\lambda \m )
\e^{b\nu}\e^{b\lambda }\e^{a\mu}\3 F^a_{\mu\nu}
\eqn\sxxx $$
This relation shows that  the constraints do form an algebra
under Poisson brackets, but the "structure constants" of this algebra
depend on the canonical variables.\nl
\chapter{The reality conditions.     \hfill\hfill}
So far we have really been dealing with {\it complex} general relativity,
and we now have to face the problem of identifying a real section.
The variable $A^a_\mu $ is complex, but we
must demand that  $\e^{a\mu}$  be real; or rather, that if they are real
initially, they remain so. Or we may be more tolerant, and let them
wander off, choosing as reality condition:
 $$  (iv)\qquad  {\tilde Q}^{\mu\nu} := \e^{a\mu}\e^{a\nu} =
real        \eqn\sli       $$
${\tilde Q}^{\mu\nu}$ is a gauge invariant tensor density of degree
2, related to the inverse 3-metric by
${\tilde Q}^{\mu\nu} = e^2 q^{\mu\nu}$.
The point is that the equations of motion \sxxix\ are completely
gauge-dependent, \ie\ they depend on  arbitrary Lagrange multipliers,
so the motion might involve a complex  gauge rotation, which in itself is
harmless, since only gauge invariant quantities like ${\tilde Q}^{\mu\nu}$
matter.  \nl
A reality condition limits the possible initial values, but  cannot be treated
like the other constraints we have met, which descend from the action.
Its consistency   can be decided by checking that, once imposed on the
initial data, it remains valid through the evolution of the system.
 By eq.\sxxix , the time derivative of  ${\tilde Q}^{\mu\nu}$ is:
$$
  {\cal L}_{\bf t}{\tilde Q}^{\mu\nu} =
{\cal L}_{\bf N}{\tilde Q}^{\mu\nu}-
   (\e^{a\mu }N^\nu +\e^{a\nu }N^\mu )\3 D_\lambda\e^{a\lambda }
- i\N \epsilon_{abc}\e^{c\lambda }(\e^{a\mu }\3 D_\lambda\e^{b\nu }-
  \e^{b\nu }\3 D_\lambda\e^{a\mu })      \eqn\slii$$
therefore we have to impose a "secondary" reality condition:
$$ (v)\qquad {\tilde P}^{\mu\nu} :=
i \epsilon_{abc}\e^{c\lambda }
(\e^{a\mu }\3 D_\lambda\e^{b\nu }-
\e^{b\nu }\3 D_\lambda\e^{a\mu }) = real  \eqn\sliii$$
Conditions (iv) and (v) together  define the real section on the
constraint surface in phase space. A useful check is to count
the  number of (real) degrees of freedom. Since $\e^{a\mu }$ and
$A^a_\mu $ are  in general complex, there are $2\cdot 2\cdot 3\cdot 3 =
36$ real dynamical variables. We have $2\cdot 3$ real constraints from
(iii), and  $1+3$ from (i) and (ii), if (as we expect)
$\tilde H$ and $\tilde F_\mu$ are real, for a total
of $10$, and $6+6$ reality conditions. So:
$$ 2\cdot {\rm n.\ of\ degrees\ of\ freedom} = 36 - 2\cdot 10 - 12 = 4$$
as it should be.\nl
However we do have to prove that $\tilde H$ and $\tilde F_\mu$ are
real, and that there are  no "tertiary" reality conditions, namely
that, once (iv) and (v) are imposed on the initial data, they  remain
valid through the evolution of the system.

 As far as I can see neither proof goes through unless
one assumes {\it non degeneracy}, namely:
$$ (*)\qquad  {\tilde Q}^{\mu\nu} = {\rm \  positive\ definite}
\eqn\sliv $$
{}From this further assumption it follows that a complex $O(3)$ gauge
transformation exists that makes $\e^{a\mu }$ real, with
$e^2 = \det (\e^{a\mu }) >0$. We can then  reconstruct the local
geometry of $\Sigma $, i.e. calculate  $e^a_\mu$ from $\e^{a\mu }$
{\it and} the Ashtekar connection in terms of $e^a_\mu$, $\tilde G^a $ and
${\tilde P}^{\mu\nu}$. In fact, using the definitions of $\tilde G^a$ and of
${\tilde P}^{\mu\nu}$, we find, after some algebra:
$$A^a_\mu =\half\epsilon_{abc}e^{b\nu}(\partial_\nu e^c_\mu
-\partial_\mu e^c_\nu -e^{c\lambda}e^d_\mu\partial_\lambda
e^d_\nu )+{1\over 2e}\epsilon_{abc}e^b_\mu \tilde G^c
+{i\over 4e^3}(2e^a_\lambda q_{\mu\rho}-
e^a_\mu q_{\lambda\rho})\tilde P^{\lambda\rho}
\eqn\slv$$
This is a very useful equation; from it we see that,
{\it if}  the conditions (iii), (iv), (v) and (*) hold, the Ashtekar connection
can be written in the form:
 $$ A^a_\mu = - \half \epsilon_{abc}q_\mu^{\ \nu}\Omega^{bc}_\nu
  +ie^{a\nu}K_{\mu\nu}
  \eqn\slvi$$
where $\Omega^{ab}_\mu$ are components of the Levi-Civita connection,
and the real, symmetric tensor
$$K_{\mu\nu} :={1\over 4e^2}
(2q_{\mu\rho}q_{\nu\lambda}-q_{\mu\nu}q_{\lambda\rho})
\tilde P^{\lambda\rho}
 \eqn\slvii$$
can be interpreted as the extrinsic curvature of $\Sigma_t$, which is
defined to be $\half {\cal L}_{\bf n}q_{\mu\nu}$.
\REF\sen{ A. Sen, J. Math. Phys. 22 (1981) 1718; Phys. Lett. 119B (1982) 89.}
\REF\witten{E. Witten, Commun. Math. Phys. 80 (1981) 725}
Eq.\slvi\ is in fact the expression one gets if one calculates $A^a_\mu$
projecting the Levi-Civita connection, and is called the "Sen connection"
[\sen ].\nl
{}From eq.\slv\ one can also prove directly that $\tilde H$ and
$\tilde F_\mu$ are real, but it is more instructive to take for $A^a_\mu$
the Sen form $A^a_\mu=q_\mu^{\ \nu}X^a_{ij}\Omega^{ij}_\nu$, for which
one obtains [\sen ][\witten ]:
$$\eqalign{
i\e^{a\mu} F^a_{\mu\nu} &=
- e\, q_\nu^{\ \rho}n^\sigma (R_{\rho\sigma}-\half g_{\rho\sigma} R) \cr
\epsilon_{abc}\e^{a\mu}\e^{b\nu} F^c_{\mu\nu} &=
-2 e^2 n^\mu n^\nu (R_{\mu\nu}-\half g_{\mu\nu} R) \cr}
    \eqn\slviii$$
Thus, provided  (iii), (iv) and (v) and (*) hold, the vector and the scalar
constraints are just the constraint part of the Einstein equations.\nl
However to assure the consistency of the scheme we still need to prove
that  the time derivative of ${\tilde P}^{\mu\nu} $ is real; which
means that we have to  express it in terms of objects which we know are
real on the real section of the constraints surface. This turns out to be a
fairly hard task, because from eq.\sxxix\ we have, with a few
simplifications:
 $$\eqalign{
&{\cal L}_{\bf t}{\tilde P}^{\mu\nu} =
 \{ {\cal H}, \tilde P^{\mu\nu}\} =
 {\cal L}_{\bf N}{\tilde
P}^{\mu\nu} -\int d^3x N^\rho \tilde G^a{\delta \tilde P^{\mu\nu} \over
\delta\e^{a\rho }}-2\N\tilde Q^{\mu\nu}\tilde H -\cropen{10pt}
&\ -2 \epsilon_{abc}\e^{a(\mu}\3 D_\lambda\e^{b\nu )}\3 D_\rho
 (\N \epsilon_{ab'c'}\e^{b'\rho }\e^{c'\lambda })
   +2 \epsilon_{abc}\e^{a\rho}\3 D_\lambda\e^{b(\nu }\3 D_\rho
(\N \epsilon_{ab'c'}\e^{b'\lambda}\e^{c'\mu )} )-\cropen{10pt}
&\ -2[\3 D_\lambda\3 D_\rho (\N \epsilon_{abc}\e^{b\rho}\e^{c(\nu })]
\epsilon_{ab'c'}\e^{b'\lambda}\e^{c'\mu )}
+2\N\tilde Q^{\lambda (\nu }\epsilon_{abc}\e^{a\mu )}\e^{b\rho}
\3 F^c_{\lambda\rho}
\cr}
 \eqn\slix $$
in which the first three terms on the R.H.S. are real or zero, but little can
be said about the rest; we cannot even be sure the expression is
diffeomorphism invariant, written in this way.  However if the metric is
non degenerate we can substitute eq.\slv\ in it, and after a {\it very}
long calculation obtain
$$\eqalign{
&{\cal L}_{\bf t}{\tilde P}^{\mu\nu}  =
 {\cal L}_{\bf N}{\tilde P}^{\mu\nu}
-\int d^3x N^\rho \tilde G^a{\delta \tilde P^{\mu\nu} \over \delta
\e^{a\rho }}-2\N\tilde Q^{\mu\nu}\tilde H -\cropen{10pt}
& -2(\tilde Q^{\mu\nu}\tilde Q^{\lambda\rho}-
\tilde Q^{\mu\rho}\tilde Q^{\lambda\nu})
\3\nabla_\lambda\3\nabla_\rho\N +
2\tilde Q^{\rho (\mu }\e^{b\nu )}\tilde G^b\3\nabla_\rho\N
-2\N  Q^{\mu\nu} \3\nabla_\lambda (\e^{b\lambda}\tilde G^b)+
\cropen{10pt}
&\ + \half\N (\tilde Q^{\mu\nu}\delta_{ab}-\e^{a\mu}\e^{b\nu})
\tilde G^a\tilde G^b  -
  i\N\epsilon_{abc}\tilde P^{\lambda (\nu }e^{a\mu )}\tilde G^b
e^{c\lambda} -2\N e^2\tilde Q^{\rho (\mu}\3 R_\rho^{\ \nu )}+
\cropen{10pt}
&\ + {\N\over 2e^2}q^{\mu\nu}(q_{\lambda\lambda'}q_{\rho\rho'}-
\half q_{\lambda\rho}q_{\lambda'\rho'})
\tilde P^{\lambda\rho}\tilde P^{\lambda'\rho'}
-{\N\over e^2}q_{\lambda\rho}
\tilde P^{\mu\lambda}\tilde P^{\nu\rho}
\cr}
 \eqn\slx$$
Here  $\3\nabla_\mu $ is the $q_{\mu\nu}$-compatible derivative on
$\Sigma_t$, $\3 R_{\mu\nu}$ the corresponding Ricci tensor.  This
unattractive expression  has the merit of being manifestly diffeomormism
invariant, and of displaying explicitely that ${\cal L}_{\bf t}{\tilde
P}^{\mu\nu} $  is indeed real if the conditions (i)...(v) {\it and} (*)   are
satisfied.

Thus, as long as $\tilde Q^{\mu\nu}$ is positive definite, the
theory is completely equivalent to Einstein's theory. When $\tilde
Q^{\mu\nu}$ is not positive definite, we cannot, in general, impose
consistently the reality conditions, and we have to make do with the
complex theory.

Again one may count the degrees of freedom: there is no reason now to
expect $\tilde H$ and $\tilde F_\mu$ to be real, so for $18$ complex
dynamical variables we have $3+1+3$ complex constraints, and therefore
$2$ complex (4 real) degrees of
freedom, and no sensible  interpretation of the $\e^{a\mu}$.

It is a pleasure to thank Abhay Ashtekar and Carlo Rovelli for some very
enlightening conversations.

\refout
\bye